\RequirePackage{fix-cm}
\RequirePackage{amsmath}
\documentclass[smallextended]{svjour3}       
\smartqed  
\usepackage[utf8]{inputenc}
\usepackage{graphicx}%
\usepackage{subcaption}%
\usepackage{siunitx}%
\usepackage{bm}%
\usepackage{mathabx}
\usepackage{booktabs}%
\usepackage[%
breaklinks=true,%
colorlinks=true,%
citecolor=blue,%
pdfauthor={Davide Amato},%
pdftitle={Lunar close encounters compete with the circumterrestrial Lidov-Kozai effect}%
]{hyperref}%
\makeatletter%
\def\cl@chapter{\@elt {theorem}}%
\makeatother%
\usepackage{cleveref}%
\usepackage{natbib}
\usepackage{todonotes}%
\usepackage{soul}%
\graphicspath{{./}}
\newcommand{\THALASSA}{\texttt{THALASSA}}

\newcommand{\figscale}{0.77}
\newcommand{\review}[1]{{#1}}
\journalname{Celestial Mechanics and Dynamical Astronomy}
\newenvironment{changemargin}[2]{%
	\begin{list}{}{%
	\setlength{\topsep}{0pt}%
	\setlength{\leftmargin}{#1}%
    \setlength{\rightmargin}{#2}%
    \setlength{\listparindent}{\parindent}%
    \setlength{\parsep}{\parskip}%
	}%
	\item[]}{\end{list}
	}

\begin{document}
 
\title{\review{Lunar close encounters compete with the circumterrestrial Lidov-Kozai effect}
\thanks{}
}
%

\subtitle{\review{The dynamical demise of \emph{Luna 3}}}


\author{Davide Amato \and
        Renu Malhotra \and \\
        Vladislav Sidorenko \and
        Aaron J. Rosengren
}


\institute{D. Amato \at
              Department of Aerospace \& Mechanical Engineering \\
              The University of Arizona \\
              Tucson, AZ, USA \\
              Tel.: +1-520-389-9688 \\
            \emph{Present address:} \\
            Colorado Center for Astrodynamics Research (CCAR) \\
            University of Colorado Boulder \\
            Boulder, CO, USA 
            \email{davide.amato@colorado.edu}
           \and
           R. Malhotra \at
              Lunar and Planetary Laboratory \\
              The University of Arizona \\
              Tucson, AZ, USA
           \and
           V. Sidorenko \at
              Keldysh Institute of Applied Mathematics, Russian Academy of Sciences \\
              Moscow, Russian Federation
           \and
           A. J. Rosengren \at
              Department of Aerospace \& Mechanical Engineering \\
              The University of Arizona \\
              Tucson, AZ, USA
}

\date{Received: date / Accepted: date}

\maketitle

\begin{abstract}
        Luna 3 (or \emph{Lunik 3} in Russian sources) was the first spacecraft to perform a flyby of the Moon.
        Launched in October 1959 on a translunar trajectory with large semi-major axis and eccentricity, it collided with the Earth in late March 1960.
        The short, 6-month dynamical lifetime has often been explained through an increase in eccentricity due to the Lidov-Kozai effect.
        \review{However, the classical Lidov-Kozai solution is only valid in the limit of small semi-major axis ratio, a condition that is satisfied only for solar (but not for lunar) perturbations.}
        We undertook a study of the dynamics of Luna 3 with the aim of assessing the principal mechanisms affecting its evolution.
        We analyze the Luna 3 trajectory by generating accurate osculating solutions, and by comparing them to integrations of singly- and doubly-averaged equations of motion in vectorial form.
        Lunar close encounters, which cannot be reproduced in an \review{averaging} approach, decisively affect the trajectory and break the doubly-averaged dynamics.
        Solar perturbations induce oscillations of intermediate period that affect the geometry of the close encounters and cause the singly-averaged and osculating inclinations to change quadrants (the orbital plane ``flips'').
        We find that the peculiar evolution of Luna 3 can only be explained by taking into account lunar close encounters and intermediate-period terms; such terms are averaged out in the Lidov-Kozai solution, which is not adequate to describe translunar or cislunar trajectories.
        Understanding the limits of the Lidov-Kozai solution is of particular significance for the motion of objects in the Earth-Moon \review{environment} and of exoplanetary systems.
\end{abstract}

\bigskip
\begin{quote}
\begin{changemargin}{10.5em}{-1.45em}
\textit{The theory of [the future motion of Luna 3] will pose a pretty problem in celestial mechanics, which the mathematicians may well choose to shirk by resorting to electronic computers.\footnote{King-Hele, op. cit., p. 685}}
\end{changemargin}

\begin{changemargin}{4.64in}{-1.5em}
(D. G. King-Hele, 1959)
\end{changemargin}
\end{quote}

\section{Introduction}
Luna 3 (COSPAR 1959-008A) was a Soviet spacecraft launched on 4 October 1959 from the Baikonur cosmodrome (at latitude \ang{46}N) that was directly inserted into a highly elliptical, translunar orbit.
The primary mission objective was to capture images of the far side of the Moon, which required the spacecraft to perform a lunar flyby (or encounter) on 6 October 1959.
The encounter was designed such that the equatorial inclination increased from \ang{55} to about \ang{80}, so that at the subsequent perigee pass the spacecraft could overfly ground stations placed in the Soviet Union to downlink scientific data~\citep{sedov1960}.\footnote{Thus, the mission achieved two firsts: the first view of the lunar far side, and the first \emph{gravity assist} maneuver.}

The trajectory of the spacecraft was reconstructed through numerical integrations by \citet{gontkovskaya1961,gontkovskaya1962,michaels1960a,michaels1960}.
In particular, \citet{gontkovskaya1961} used a variable-step Runge-Kutta 4(5) solver to integrate the equations of motion with Encke's method.
They showed that the eccentricity monotonically increased from the launch epoch onward, leading to an eventual Earth collision on 30 March 1960, after 11 revolutions had been completed.
The simulations also revealed that the spacecraft performed a second lunar flyby on 24 January 1960 \review{that} considerably affected the semi-major axis and inclination.
Further analyses of the trajectory demonstrated that the increase in eccentricity was ascribable to the effects of lunar and solar perturbations, each of whose \review{influence} on the orbital elements were comparable in magnitude~\citep{gontkovskaya1962}.
The fact that a spacecraft re-entry can be caused purely by gravitational interactions (rather than by energy dissipation due to atmospheric drag) is nowadays well known, and it has even been proposed as a natural mechanism for end-of-life disposal \citep{rosengren_galileo_2017,namazyfard_computational_2019,Gkolias2019,Skoulidou2019}.
However, the novelty of this idea commanded attention in scientific exchanges of the 1950s~\citep{upton_lunar_1959,kozai_explorer_1959}.
For instance, \citet{sedov1960} notes that ``as a result of perturbation due to the sun [\ldots] only satellites with orbits of certain types can ``survive'' for a long time,'' and \citet[p. 101]{beletsky2001} even reports that ``a respected scientist [\ldots] declared that [Luna 3] will fly in its orbit forever'' in a public lecture.

Until the advent of artificial satellites, studies of the circular, restricted three-body problem (CR3BP) were largely limited to orbits of natural Solar System bodies, which are nearly coplanar and have only moderate eccentricity.
Asteroids and comets were outstanding exceptions, as many of their orbits were found to be highly eccentric and inclined at large angles with respect to the ecliptic.
The necessities of the Space Age rekindled the interest in highly eccentric and inclined orbits~\citep{king-hele_orbits_1959,margerison_way_1958,egorov_certain_1958}, the importance of which was underscored by the unexpected re-entries of Explorer VI and Luna 3; the latter may have motivated Soviet dynamicist M. L. Lidov to study the \review{evolution} of highly inclined orbits in the CR3BP.
In \citet{lidov1961,lidov1962}, he developed a solution for highly inclined orbits in the doubly-averaged CR3BP with Hill's approximation, in which the perturbing function expansion is truncated to the second (or \review{\emph{quadrupolar}}) order in the ratio of semi-major axes. 
Having likely been inspired by Lidov's work, \citet{kozai1962} derived the same solution using von Zeipel's method to doubly-average the CR3BP Hamiltonian.
Since the twice-averaged Hamiltonian is reduced to a single degree of freedom, the problem is integrable and admits three integrals of motion.
The orbital energy and the vertical component of the angular momentum are constant, and the eccentricity and inclination undergo coupled oscillations (the so-called \emph{Lidov-Kozai oscillations}).
The global behavior of the solution can be appreciated by plotting iso-potential curves on the $(1-e^2,\omega)$ plane in \emph{Lidov-Kozai diagrams}.
In the last few decades, research on extensions and applications of the Lidov-Kozai theory has seen considerable activity \citep[fig. 29]{ito_lidov-kozai_2019}.
\citet{naoz_hot_2011,katz_long-term_2011,lithwick_eccentric_2011} expounded on the consequences of introducing the third-order (or \emph{octupolar}) term in the perturbing function.
The third-order solution shows that the amplitude of the Lidov-Kozai cycles is modulated on a slow timescale, and ``flips'' of the orbital plane, in which $\cos i$ changes sign, become possible.
Since the octupolar term is directly proportional to the eccentricity of the perturber, the solution has been called the ``Eccentric Kozai-Lidov effect'' (EKL).
The EKL has been recently shown by one of the authors to be equivalent to a resonance phenomenon \citep{sidorenko_eccentric_2017}.
A complementary line of studies has been devoted to the further relaxation of the small semi-major axis hypothesis.
It turns out that even when the orbits of the test particle and of the perturber intersect, it is still possible to doubly-average the Hamiltonian and obtain piecewise smooth solutions under the hypothesis that the semi-major axis is piecewise constant \citep{lidov_analysis_1974,gronchi_averaging_1998}; the librations exhibited by these solutions are typical of Lidov-Kozai dynamics \citep{gronchi_stable_1999}.
A general classification of orbits in the doubly-averaged CR3BP for all values of $\alpha$ was carried out by \citet{vashkovyak_evolution_1981}. 
Moreover, \citet{shevchenko2017} presented a comprehensive review of the dynamical fundamentals of the Lidov-Kozai theory, including its generalizations, and its application to stellar and exoplanetary systems.
\review{A striking consequence of the classical Lidov-Kozai theory is that, if the Earth's non-spherical gravitational potential were less significant, the eccentricity of highly inclined low-Earth orbits would increase up to collisional values.
This is not the case because the precession of the argument of perigee caused by the Earth's $J_2$ (quadrupole) potential suppresses the Lidov-Kozai effect, and this mechanism is attributable to a coincidental mathematical property of the $J_2$ potential \citep{tremaine_why_2014}.}
During the writing of this article, we also became aware of a recent, in-depth analysis of the classical Lidov-Kozai theory, its extensions, and its relation to antecedent studies by von Zeipel~\citep{ito_lidov-kozai_2019}.
We highly recommend this work to the reader interested in the details and history of the topic.




The evolution of Luna 3 and its eventual re-entry have been explicitly attributed to the Lidov-Kozai effect (\citeauthor{morbidelli2002} \citeyear{morbidelli2002}, p. 159, \citeauthor{shevchenko2017} \citeyear{shevchenko2017}, sec. 1.4, \citeauthor{batygin_yoshihide_2018} \citeyear{batygin_yoshihide_2018}).
However, the foundational assumption of small semi-major axis in the classical Lidov-Kozai theory is completely violated by the orbit of Luna 3 for lunar perturbations.
Jointly with a high eccentricity, the large semi-major axis also resulted in the orbit intersecting that of the Moon, in which case the Legendre expansion for the inner hierarchical problem diverges and the classical Lidov-Kozai theory is not applicable.
In light of these considerations, the characterization of the trajectory of Luna 3 as an instance of the Lidov-Kozai effect is dubious.

The aim of this article is to carry out a detailed dynamical study of the trajectory of Luna 3, and to assess to what extent it was influenced by Lidov-Kozai dynamics.
Although this research question may superficially appear as a historical curiosity, the study has important implications for future cislunar and translunar missions.
Experience from ongoing missions shows that in-depth knowledge of cislunar dynamics carries a substantial impact on mission design.
For instance, the initially chaotic orbit of the Interstellar Boundary EXplorer (IBEX) was placed in a stable equilibrium of the $3:1$ lunar mean-motion resonance (MMR) through a phasing maneuver that extended its lifetime by several decades \citep{carrico2011,mccomas2011}.
In a similar fashion, the Transiting Exoplanet Survey Satellite (TESS) reached a stable orbit in a $2:1$ lunar MMR through a flyby of the Moon~\citep{gangestad_high_2014}.
With renewed interest in translunar missions stirred by the NASA \review{Artemis} program \citep{dunbar_nasa_2019}, we expect that knowledge of the dynamics of the translunar region will become even more relevant.

The article is organized as follows.
A brief theoretical outline of the Lidov-Kozai effect, which is a prerequisite for the interpretation of the results, is given in \cref{sec:LKE_theory}.
We reconstruct the trajectory of Luna 3 from published ephemerides and analyze its sensitivity to perturbations and uncertainties in the initial conditions in \cref{sec:nominal_trajectory}.
Using these considerations, we compare doubly-averaged, singly-averaged, and osculating evolutions of the spacecraft in \cref{sec:oscVsAvg}, and we assess whether the Lidov-Kozai theory gives meaningful predictions for the case of Luna 3 in \cref{sec:Luna3_LKE}.
Finally, we discuss the achieved results in \cref{sec:concl}.



\section{The Lidov-Kozai effect}
\label{sec:LKE_theory}
As to properly situate Luna 3 in the context of the restricted three-body problem, it is necessary to briefly introduce the foundations of the Lidov-Kozai effect.
For additional details, the reader is addressed to \citet{lidov1961,lidov1962,kozai1962,malhotra2012,shevchenko2017,ito_lidov-kozai_2019}.

Consider an orbit of semi-major axis $a$ and mean motion $n$ of a massless spacecraft around a central mass perturbed by a distant massive body; for instance, the orbit of an artificial satellite around the Earth that is perturbed by the Sun.
The perturbing body moves on a circular orbit with mean motion $n'$, in accordance with the hypotheses of the CR3BP, and we take its orbital plane as a reference for the measurement of all angles.
The corresponding perturbing function is expressed as~\citep{murray2000}
\begin{equation}
  \mathcal{R} = \mu' \left( \frac{1}{\lVert \bm{r} - \bm{r}' \rVert} - 
  \frac{\bm{r} \cdot \bm{r}'}{\lVert \bm{r}' \rVert^3} \right).
  \label{eq:R_coord}
\end{equation}
Assuming that $r < r'$, the first term in the above equation above can be expanded in Legendre polynomials as
\begin{equation}
  \frac{1}{\lVert \bm{r} - \bm{r}' \rVert} = \sum_{l=0}^{\infty} \left( \frac{r}{r'} \right)^l P_l\left( \cos\psi \right),\label{eq:oneOverD_expansion}
\end{equation}
where $\cos\psi = (\bm{r} \cdot \bm{r}')/rr'$.
By plugging \cref{eq:oneOverD_expansion} into \cref{eq:R_coord} and neglecting a nonessential constant, we obtain the classical Legendre expansion for the perturbing function,
\begin{equation}
  \mathcal{R} = \frac{\mu'}{r'} \sum_{l=2}^{\infty} \left( \frac{r}{r'} \right)^l P_l\left( \cos\psi \right).
  \label{eq:R_expansion}
\end{equation}
In the quadrupolar (or Hill's) approximation, the expansion is truncated at $l = 2$, giving:
\begin{equation}
  \mathcal{R} \approx \frac{\mu'}{2r'} \left( \frac{r}{r'} \right)^2 \left( 3 \cos^2 \psi - 1 \right).
  \label{eq:R_quadrup}
\end{equation}
In the absence of MMRs, the disturbing function can be averaged over the mean anomaly of the spacecraft as
\begin{equation}
  \overline{\mathcal{R}} = \frac{1}{2\pi} \int_{0}^{2\pi} \mathcal{R} \, \mathrm{d}M.
\end{equation}
Substitution of $\overline{\mathcal{R}}$ into the Lagrange planetary equations leads to the singly-averaged equations of motion, admitting two integrals~\citep{vashkovyak_particular_2005}:
\begin{align}
a &= \text{const.} \\
\overline{\mathcal{R}} + \nu \sqrt{ 1 - e^2 } \cos i &= \text{const.},
\end{align}
where the first integral arises due to the fact that $\overline{\mathcal{R}}$ does not depend on the mean anomaly of the satellite, and the second is obtainable through a change of variables that removes the time dependence.
The quantity
\begin{equation}
  \nu = \frac{16}{3} \frac{\mu}{\mu'} \frac{n'}{n} \frac{(a')^3}{a^3}
\end{equation}
is a dimensionless parameter that is approximately equal to $16n/3n'$ for $\mu/\mu' \ll 1$.

The singly-averaged Hamiltonian can be written as
\begin{equation}
\overline{\mathcal{H}} = -\frac{\mu}{2a} - \overline{\mathcal{R}}.
\end{equation} 
After a subsequent averaging over the mean anomaly of the perturber, the doubly-averaged perturbing function is given by~\citep{malhotra2012} 
\begin{equation}
  \overline{\overline{\mathcal{R}}} = \frac{\mu' a^2}{8 a'^3} \left[ 2 + 3e^2 - 3 \left( 1 - e^2 + 5e^2 \sin^2 \omega \right) \sin^2 i \right],
  \label{eq:LKH_classical}
\end{equation}
where a superfluous constant has \review{again been} neglected.
Since $\overline{\overline{\mathcal{R}}}$ does not depend on the mean longitude $l$ and on the longitude of the ascending node $\Omega$, the conjugate Delaunay momenta $L = \sqrt{a \left(\mu + \mu' \right)}$ and $H = L \sqrt{1 - e^2} \cos i$ are constants.
In addition, the doubly-averaged Hamiltonian does not depend explicitly on time or on the mean anomaly and it is also constant.
Thus, the doubly-averaged problem is completely integrable and possesses three integrals,
\begin{align}
  c_0 &= a = \text{const.} \\
  c_1 &= \left(1 - e^2 \right) \cos^2 i = \text{const.} \label{eq:c1}\\
  c_2 &= e^2 \left(\frac{2}{5} - \sin^2 i \sin^2 \omega \right) = \text{const.} \label{eq:c2}
\end{align}
Substituting the value of $\sin^2 i$ from \cref{eq:c1} into \cref{eq:c2} results in an alternate expression for $c_2$ as a function of $(c_1, e, \omega)$,
\begin{equation}
  c_2\left( e, c_1, \omega \right) = e^2 \left[ \frac{2}{5} - \left( 1 - \frac{c_1}{1 - e^2} \right) \sin^2 \omega \right].
  \label{eq:c2_ec1w}
\end{equation}
Solutions can be visualized in the $(\omega,X \triangleq \sqrt{1 - e^2})$ plane as isolines of the $c_2$ integral, for a given $c_1$.
These flow plots, as originally presented by \review{\citet{lidov_approximated_1963} and \citet{kozai1962}}, are called \emph{Kozai} or \emph{Lidov-Kozai} diagrams.

By writing either the Lagrange planetary equations or the rates of change of the Delaunay variables one recognizes that there exist stationary solutions such that $\dot{\omega} = 0$ above a critical inclination $i_\mathrm{crit}$.
In the limit of small semi-major axis, $\alpha \triangleq (a/a') \ll 1$, the critical inclination is $i_\mathrm{crit} \approx \SI{39.2}{\degree}$.
Close to these fixed points, the argument of perigee librates on a slow timescale.
The libration of the perigee excites coupled oscillations in eccentricity and inclination, and the integrals $c_1$ and $c_2$ are conserved according to \cref{eq:c1,eq:c2}.
In particular, taking the derivative of \cref{eq:c1} with respect to time one sees that the eccentricity and inclination are coupled through
\begin{equation}
  \tan i \frac{\mathrm{d}i}{\mathrm{d}t} = - \frac{e}{1 - e^2} \frac{\mathrm{d}e}{\mathrm{d}t}.
  \label{eq:i_e_coupling}
\end{equation}

\subsection{The four-body Hamiltonian in the quadrupolar approximation}
\label{sec:4bodyHamiltonian}
Let us now consider a restricted four-body problem with the Earth, Sun and the Moon as massive bodies on fixed orbits, and the spacecraft in a geocentric orbit as the massless body.
We treat the Sun and the Moon as two massive perturbers on \review{non-intersecting hierarchical orbits that do not interact gravitationally} (an acceptable approximation for short timescales).
Then the Hamiltonian, averaged over the mean anomalies of the \review{spacecraft}, the Sun and the Moon, is
\begin{equation}
  \overline{\overline{\mathcal{H}}} = -\frac{\mu}{2a} - \left( \overline{\overline{\mathcal{R}}}_\Sun + \overline{\overline{\mathcal{R}}}_\Moon \right).
\end{equation}
The expressions for the lunar and solar perturbing functions $\overline{\overline{\mathcal{R}}}_\Sun$ and $\overline{\overline{\mathcal{R}}}_\Moon$ are obtained by considering the orbital elements in Jacobi coordinates\footnote{That is to say that the orbital elements of the Moon and of the spacecraft are referred to the Earth's center, and those of the Sun are referred to the Earth-Moon barycenter.} in \cref{eq:LKH_classical}, and by substituting the respective values for $\mu'$ and $a'$.
The averaging operation is performed three times: once over the period of the spacecraft, and once over the period of each of the perturbers.
The inclination and argument of perigee in \cref{eq:LKH_classical} are measured with respect to a reference frame centered in the Earth and with the perturber's orbital plane as its fundamental plane.
Assuming that the orbital plane of the Moon coincides with the ecliptic, the inclination and argument of perigee are the same whether they are measured with respect to the orbital plane of the Moon or of the Sun, and the term in brackets in \cref{eq:LKH_classical} is identical in $\mathcal{R}_\Sun$ and $\mathcal{R}_\Moon$.
The approximation holds quite well in reality, as the orbital plane of the Moon is only inclined by \ang{5;15;} with respect to the ecliptic. 
Therefore,
\begin{equation}
  \overline{\overline{\mathcal{R}}} = \overline{\overline{\mathcal{R}}}_\Sun + \overline{\overline{\mathcal{R}}}_\Moon = \frac{a^2}{8} \left( \frac{\mu_\Sun}{a_\Sun^3} + \frac{\mu_\Moon}{a_\Moon^3} \right) \left[ 2 + 3e^2 - 3 \left( 1 - e^2 + 5e^2 \sin^2 \omega \right) \sin^2 i \right].
  \label{eq:LKH_4body}
\end{equation}
The equation above shows that if the two perturbers are in the same orbital plane, the four-body perturbing function has the same form as in the three-body case.
Importantly, the four-body perturbing function $\overline{\overline{\mathcal{R}}}$ is still axisymmetric, and thus the normal vertical component of the angular momentum $c_1$ is still conserved.
Henceforth, all the developments from the classical Lidov-Kozai theory for the CR3BP carry over to the coplanar four-body problem.
The only change stemming from adding the Moon as a perturber to a Earth-spacecraft-Sun three-body problem is an increase of the factor characterizing the strength of the perturbing function from $\mu_\Sun/a_\Sun^3$ to $\left( \mu_\Sun/a_\Sun^3 + \mu_\Moon/a_\Moon^3 \right)$.

We conclude this section with a remark on the significance of higher order terms for high-altitude Earth satellite orbits.
The solar perturbing function is well approximated by the quadrupolar term since the spacecraft's semi-major axis is always small compared to that of the Sun.
On the other hand, the semi-major axis ratio with respect to the Moon can be large for high-altitude satellites; for instance, in the case of Luna 3, $a/a_\Moon = 0.689$.
However, not all higher order terms of the lunar perturbing function are important.
Odd terms of the lunar perturbing function are proportional to the eccentricity of the Moon and can therefore be neglected since $e_\Moon \approx 0.0549$.
Moreover, the octupolar timescale is significantly larger than the lifetime of Luna 3.
We will not consider extensions of the Lidov-Kozai theory to octupolar order in this work.

\section{Nominal trajectory and minimal physical model}
\label{sec:nominal_trajectory}
\begin{table}[t]
  \centering
\caption{\review{Initial osculating orbital elements for the numerical integration at the epoch 15 October 1959, 15:00 UTC (MJD 36856.625). The left column and center columns give the orbital elements in the equator and equinox frames of Besselian epoch 1959, and J2000. The right column shows the orbital elements after the correction to the mean anomaly described in \cref{sec:nominal_trajectory}, which are the nominal initial conditions used in the study.\label{tab:ICs}}}
  \[
  \begin{array}{c*3S[tight-spacing=false,table-format=6.7]s}
      \toprule
      {}              & {\text{B1959}} & {\text{J2000}} & {\text{J2000, corrected}} & {}          \\
      \midrule                                                                                       
      a               & 264557.08      & 264557.08      & 264557.08                 & \kilo\metre \\
      e               & 0.8217846      & 0.8217846      & 0.8217846                 &             \\
      i               & 79.657         & 79.872         & 79.872                    & \degree     \\
      \Omega          & 251.613        & 251.627        & 251.627                   & \degree     \\
      \omega          & 181.882        & 181.806        & 181.806                   & \degree     \\
      M               & 203.214        & 203.214        & 289.883                   & \degree     \\
      \bottomrule
  \end{array}
  \]
\end{table}
Several works reconstructed the trajectory of Luna 3 from either Soviet or Western observations of the spacecraft~\citep{gontkovskaya1961,michaels1960,michaels1960a,gontkovskaya1962,sedov1960}.
\review{In particular, \citet[section 3, GC61 from here on]{gontkovskaya1961} published the full set of osculating elements with respect to the equator and equinox of epoch \num{1959.0} on 15 October, 1959, 15:00 UTC.}\footnote{\review{GC61 seems to be the only original public source of ephemerides for Luna 3.} No Two-Line Elements (TLEs) \review{for the spacecraft} are present in the US Space Object Catalog. The mission predates the establishment of the catalog, and no TLEs seem to have been derived or added to the catalog \emph{a posteriori}. Moreover, no Luna 3 ephemerides are present in the JPL HORIZONS system \citep{park_horizons_2019}. \review{State vectors identical to those obtained by numerical integration in GC61 are given by \citet[p. 4]{king-hele_rae_1987}.}}
We took this set of osculating elements as initial conditions for a numerical integration using the numerical orbit propagation code \THALASSA{} \citep{amato2019}, considering gravitational perturbations by the Sun and the Moon exclusively, whose positions are provided by the DE431 ephemerides \citep{folkner_planetary_2014}.
The initial epoch is after the first lunar flyby of 6 October 1959, therefore no sudden changes in the orbital elements due to this encounter will appear in the reconstructed trajectory.

To rigorously reproduce \review{the trajectory, we start from the orbital elements on p. 92 of GC61, which are expressed in the true equator and equinox reference frame of 1959.0 (left column of \cref{tab:ICs}).}
\review{We reduce the orbital elements to the mean equator and equinox of J2000 (center column of \cref{tab:ICs}), which is the reference frame used by \THALASSA{}, according to the IAU 2006A conventions for precession and nutation~\citep[section 5]{petit_iers_2010}.}
\review{Then, we apply a correction to the initial mean anomaly by back-propagating its value from 18 October 1959 16:48 UTC (first row, second column of Table 4 of GC61) to 15 October 15:00 UTC according to Keplerian motion.
The orbital elements after the change of reference frame and the correction of the initial mean anomaly are in the right column of \cref{tab:ICs}, and are the nominal osculating initial conditions for the remainder of this study.}

The history of the orbital elements starting from the \review{nominal} initial conditions is shown in~\cref{fig:history}, where the black line is the output from our numerical integration and the circles correspond to the ephemerides in Table 4 of GC61.
The numerical integration starting from the modified initial mean anomaly is in excellent agreement with GC61; thus we consider the initial conditions in \cref{tab:ICs} as the nominal for the rest of the work.
The necessity of proper phasing in reproducing the trajectory already suggests that singly- and doubly-averaging methods, which remove the dependence on the mean anomaly entirely, might not be able to reproduce the evolution.

The evolution of the radius of perigee and of the selenocentric distance is shown in \cref{fig:rp_lundist}, where we also plot the data from GC61 for the radius of perigee.
Two lunar close encounters take place on 24 January 1960 and 18 March 1960 with periselenes of \SI{62280}{\kilo\meter} and \SI{95630}{\kilo\meter} respectively, the importance of which will become apparent later.
Only the first close encounter is reported in the literature~\citep[GC61,][]{michaels1960}, although the possibility of subsequent encounters was already recognized by \citet{king-hele_orbits_1959}.
The eccentricity increases as a result of the geometry of each encounter, which is readily seen by comparing~\cref{fig:rp} against the top right panel of~\cref{fig:history}.

\subsection{Sensitivity to the initial conditions and to the dynamical model}
\begin{figure}
\centering
\includegraphics[width=0.8\linewidth]{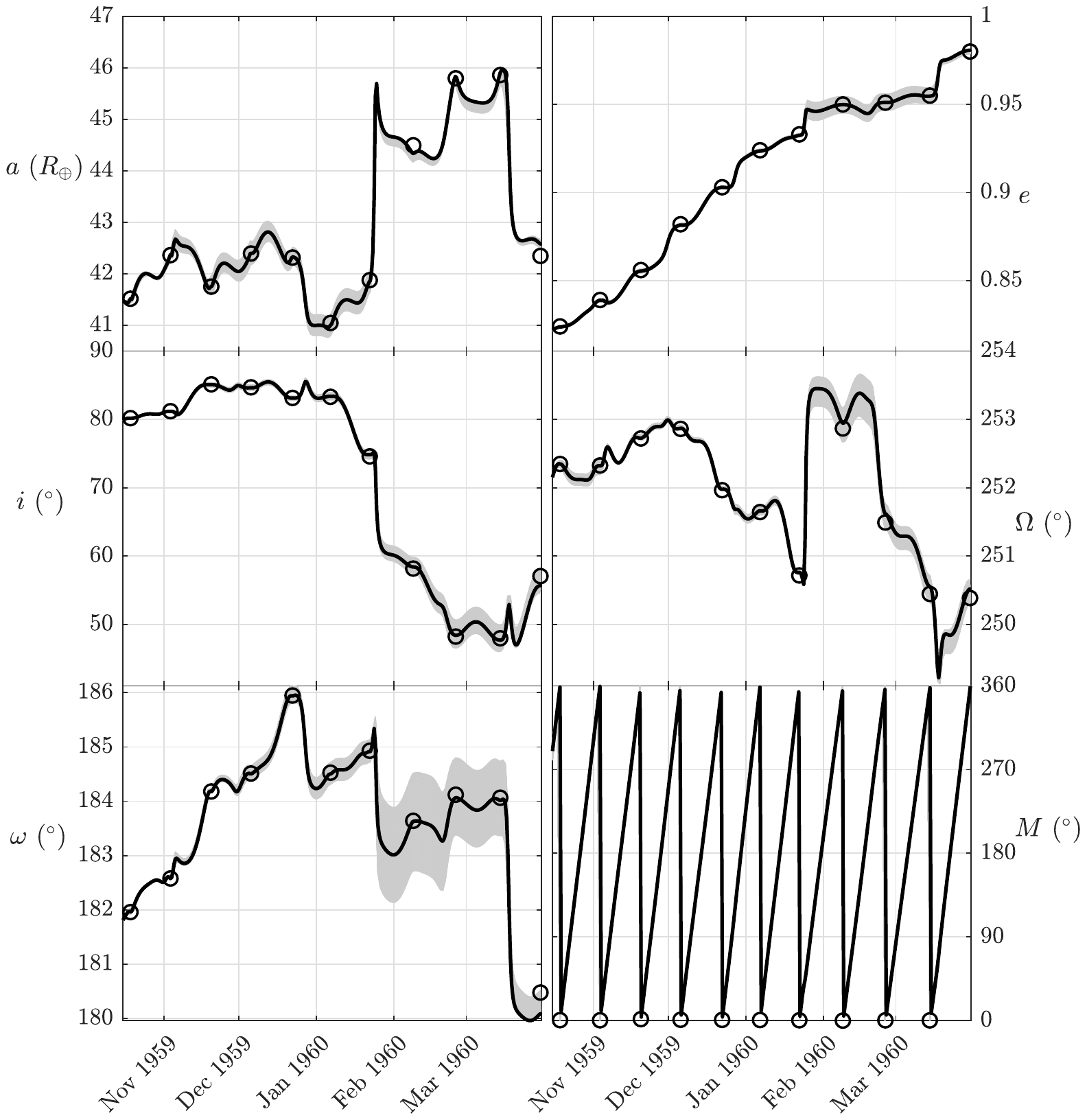}
\caption{Evolution of the orbital elements in the J2000 reference frame. Black lines correspond to the numerical propagations of the nominal initial conditions in~\cref{tab:ICs}, grey lines correspond to an ensemble of propagations with initial mean anomaly varying between \ang{279.9;;} and \ang{299.9;;}, and circles correspond to the orbital elements in Table 4 of \citet{gontkovskaya1961}. The semi-major axis is expressed in Earth radii. \label{fig:history}}
\end{figure}

\begin{figure}[t]
\centering
\begin{subfigure}[b]{.4\linewidth}
   \includegraphics[scale=\figscale]{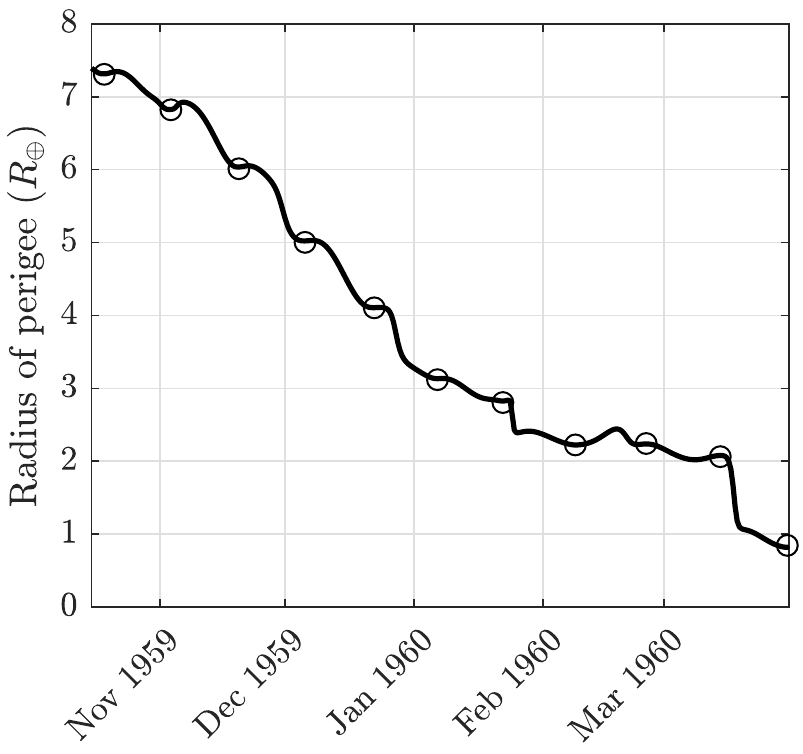}
   \caption{\label{fig:rp}}
\end{subfigure}\hspace{0.3in}%
\begin{subfigure}[b]{.4\linewidth}
   \includegraphics[scale=\figscale]{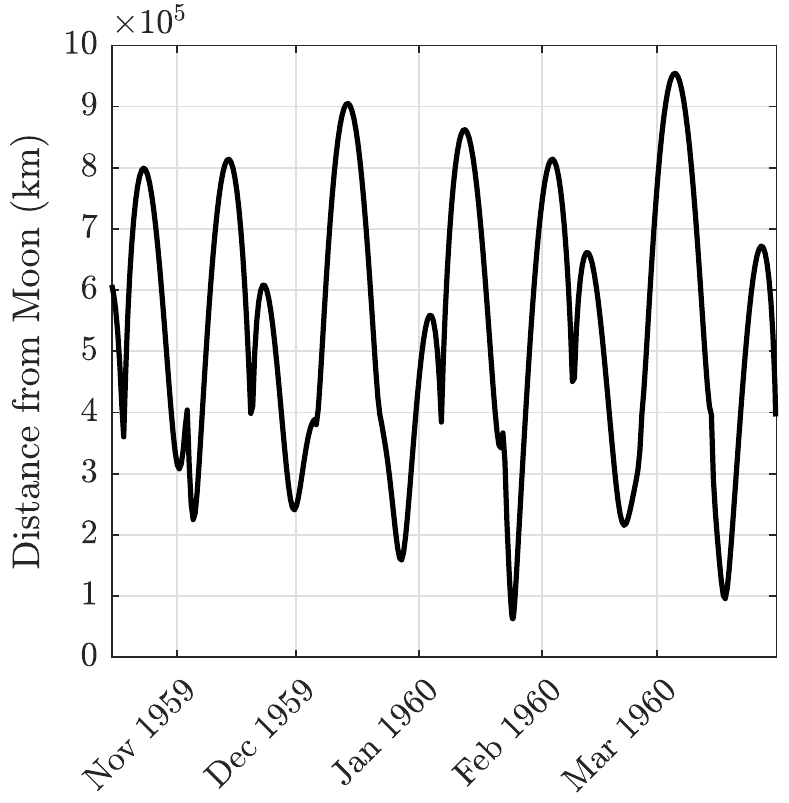}
   \caption{\label{fig:lunDist}}
\end{subfigure}
\caption{Evolutions of the radius of perigee and of the distance from the Moon from the numerical propagation of the nominal initial conditions in~\cref{tab:ICs}. Black circles in the left panel correspond to Table 4 of \citet{gontkovskaya1961}.\label{fig:rp_lundist}}
\end{figure}

We propagate an ensemble of trajectories with initial orbital elements in \cref{tab:ICs}, but with the mean anomaly varying in a $\pm \ang{10;;}$ interval around the nominal value of \ang{289.883}, with a spacing of \ang{0.1;;}; these are displayed in gray in~\cref{fig:history}.
We vary the mean anomaly since it is the variable affected by the largest uncertainty.
The ensemble of trajectories does not diverge significantly from the nominal, indicating that the dynamics are not highly sensitive to small changes in the initial (recovered) mean anomaly.
All of the trajectories in the ensemble collide with the Earth at the end of March 1960, in accordance with the re-entry dates reported by GC61 and \citet{king-hele_rae_1987}.
Other re-entry dates reported in the literature are 8 March 1960 \citep{michaels1960,michaels1960a} and 20 April 1960 \citep{williams_luna_2019}, but specific knowledge on how these re-entry dates were obtained (such as initial conditions of numerical integrations) is lacking.

We examined the sensitivity of the trajectory to perturbations other than lunisolar by propagating the nominal initial conditions with a comprehensive dynamical model including solar radiation pressure, atmospheric drag, a $15 \times 15$ Earth non-spherical gravitational potential, and a constant Earth rotation rate, besides lunisolar perturbations.
The area-to-mass ratio was \SI{4e-3}{\square\meter\per\kilo\gram}, and the drag and reflectivity coefficients were \num{2.2} and \num{1.5}, respectively.
Except for modest spikes in the semi-major axis caused by short-periodic perturbations from $J_2$, the trajectory does not differ significantly from that with lunisolar perturbations only, and we omit showing the corresponding evolution of the orbital elements for brevity.
The test confirms that lunisolar perturbations are the main driver of the dynamics, thus we will omit all other perturbations in the following.

\section{Osculating and averaged dynamics in the Earth-spacecraft-Moon-Sun 4-body problem}
\label{sec:oscVsAvg}
In this section, we \review{compare} the osculating evolution to trajectories that are singly-averaged (with respect to the orbital period) and doubly-averaged (with respect to the periods of the Sun and/or of the Moon).
We consider trajectories that are either singly-averaged with respect to the spacecraft's orbital period, or doubly-averaged with respect to the periods of the Sun and/or the Moon.
All the averaged solutions are obtained by numerical integration of the equations of motion for mean Milankovitch elements \citep{rosengren2014}, in which the perturbing functions are expressed to the quadrupole order.
\review{The integration of each of the trajectories is terminated as soon as a collision with the Earth is detected. In \THALASSA{}, a collision takes place as soon as the geocentric distance $r < R_\Earth + \SI{120}{\kilo\meter}$, where $R_\Earth = \SI{6378.136}{\kilo\meter}$ is the radius of the Earth. In the averaged solutions, a collision is assumed to take place when the mean radius of perigee is less than $R_\Earth$. This leads to a slight underestimation of the spacecraft lifetime that does not meaningfully affect the results.}
To separate the effects of the Sun and the Moon, we first take into account the case in which the Sun is the only perturber, and then we add the Moon to the dynamical model.

\subsection{Numerical osculating-to-mean transformations}
\label{sec:ICs_NumAvg}
The singly-averaged initial conditions $\bm{x}_\mathrm{SA}(t_0)$ are obtained through a numerical quadrature of the osculating trajectory,
\begin{equation}
  \bm{x}_\mathrm{SA}\left( t_0 \right) = \frac{1}{T} \int_{t_0 - T/2}^{t_0 + T/2} \bm{x}\left( t \right) \mathrm{d}t, \label{eq:SA_ICs}
\end{equation}
where $T$ is the orbital period of the spacecraft at $t_0$, the epoch corresponding to 15 October 1959, 15:00 UTC.
For the case in which the Sun is the only perturber, the doubly-averaged initial conditions are
\begin{equation}
  \bm{x}_\mathrm{DA}\left( t_0 \right) = \frac{1}{T_\Sun} \int_{t_0 - T_\Sun/2}^{t_0 + T_\Sun/2} \bm{x}_\mathrm{SA}\left( t \right) \mathrm{d}t, \label{eq:DA_SunOnly}
\end{equation}
where $T_\Sun$ is the Sun's period, otherwise we also have to perform an average over the lunar period $T_\Moon$,
\begin{equation}
  \bm{x}_\mathrm{DA}\left( t_0 \right) = \frac{1}{T_\Sun T_\Moon} \int_{t_0 - T_\Sun/2}^{t_0 + T_\Sun/2} \left[ \int_{t - T_\Moon/2}^{t + T_\Moon/2} \bm{x}_\mathrm{SA}\left( \tau \right) \mathrm{d} \tau \right] \mathrm{d}t. \label{eq:DA_SunMoon}
\end{equation}
In \cref{eq:DA_SunOnly,eq:DA_SunMoon}, the singly-averaged trajectory $\bm{x}_\mathrm{SA}(t)$ is numerically integrated from the initial conditions computed through \cref{eq:SA_ICs}.
All the integrals are computed through numerical quadratures.

\subsection{Superimposing averaged and osculating evolutions}
\begin{figure}[tp]
  \begin{subfigure}[b]{.5\linewidth}
    \centering
    \includegraphics[scale=0.79]{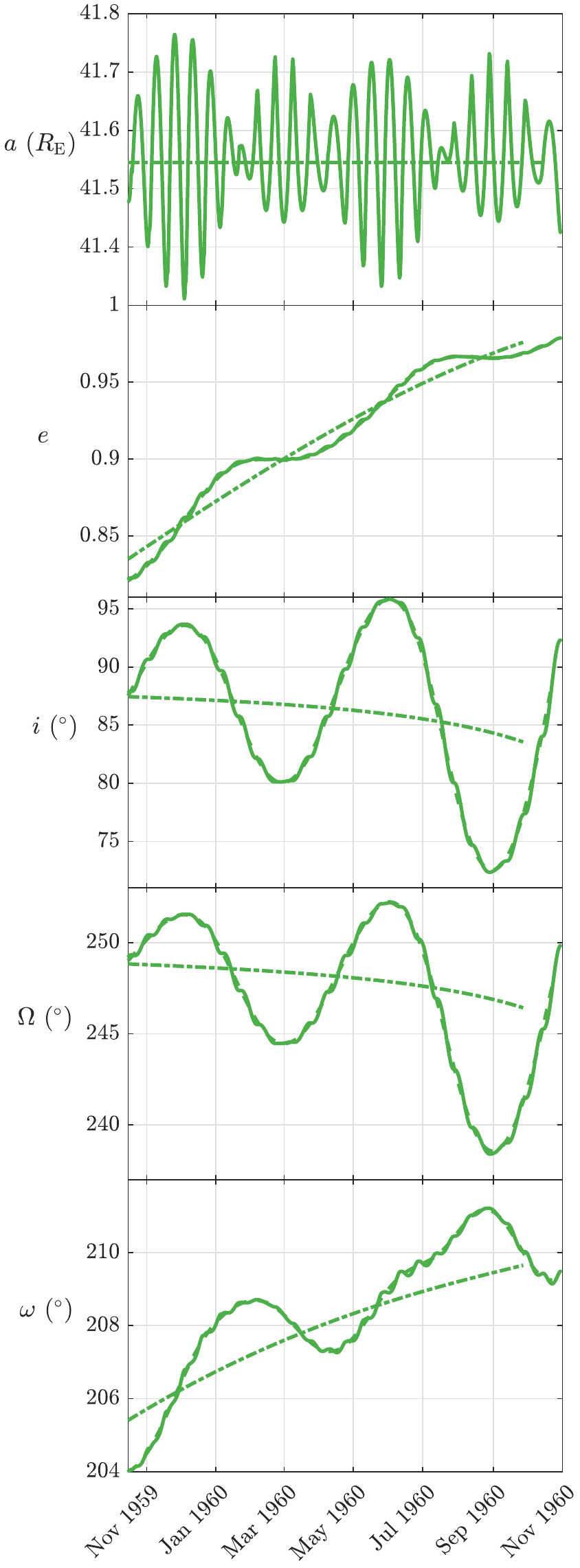}
    \caption{\review{Sun only}\label{fig:Sun_1y}}
  \end{subfigure}%
  \begin{subfigure}[b]{.5\linewidth}
    \centering
    \includegraphics[scale=0.79]{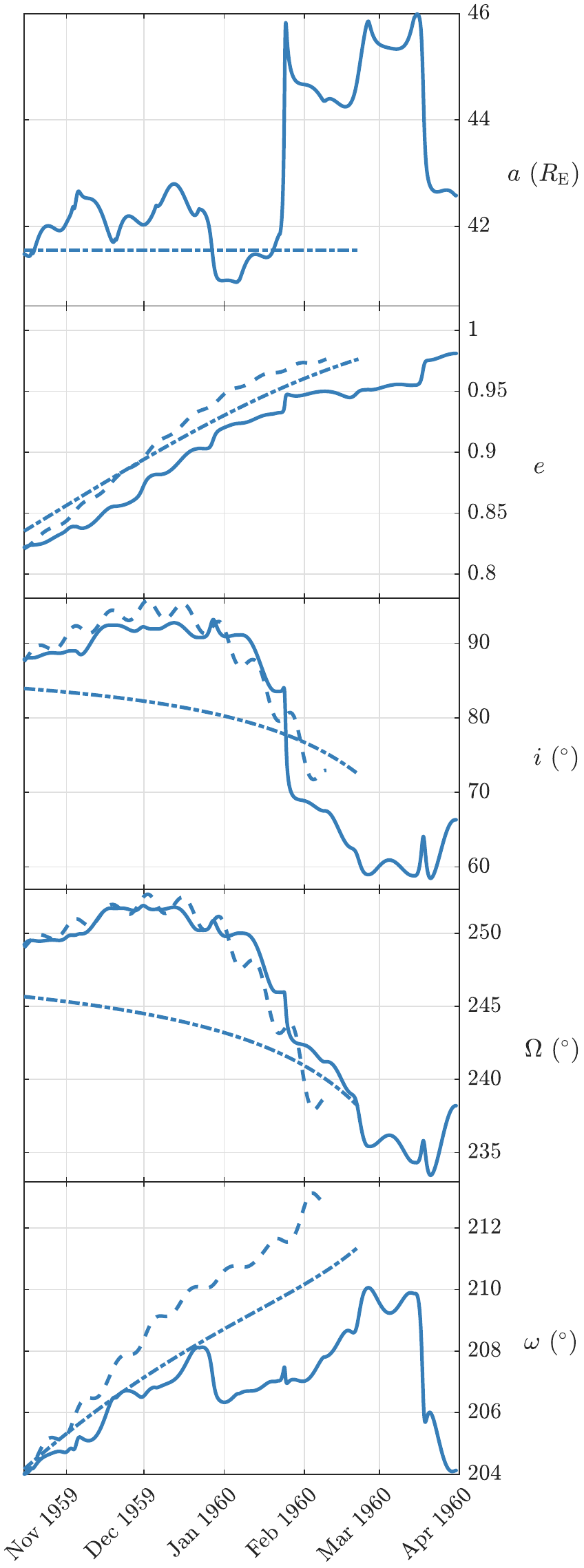}
    \caption{\review{Sun, Moon}\label{fig:SunMoon_1y}}
  \end{subfigure}
  \caption{\review{Evolution of the slow orbital elements starting from 15 October 1959 until collision with the Earth is detected, with angles measured in the J2000 ecliptic-equinox reference frame. Either solar perturbations exclusively (left panels) or both \review{solar and lunar} perturbations (right panels) are considered in the dynamical model. Continuous lines denote osculating elements, dashed lines denote singly-averaged elements, and dot-dashed lines denote doubly-averaged elements.}}
\end{figure}

In order to compare our simulations to the expected behaviors from the Lidov-Kozai theory, in which all the elements are referred to the orbital plane of the perturbing body, \emph{we measure all angles with respect to the ecliptic} in the following developments.

\Cref{fig:Sun_1y} displays the osculating, singly-averaged and doubly-averaged elements $(a,e,i,\Omega,\omega)$ as a function of time with solar perturbations exclusively, with the singly- and doubly-averaged initial conditions having been derived following \cref{sec:ICs_NumAvg}.
The trajectory follows a regular trend that is well described by the averaged equations.
The osculating semi-major axis oscillates around the initial value with a frequency corresponding to the mean motion, without displaying any secular behavior.
Since the doubly-averaged inclination is decreasing and always less than \SI{90}{\degree}, $\tan i > 0$ and the doubly-averaged eccentricity increases according to \cref{eq:i_e_coupling}.
%

The averaging with respect to the Sun's orbital period removes oscillations of periods that are larger than that of the spacecraft, but smaller than the secular timescale.
As seen from \cref{fig:Sun_1y}, these \emph{intermediate-period} oscillations~\citep{nie_semi-analytical_2019} are significant compared to the variation of the doubly-averaged orbital elements.
In particular, their magnitude is sufficient to flip the direction of motion between direct and retrograde five times before the spacecraft eventually collides with the Earth.
%
%
%
The evolution of the orbital elements in the presence of both lunar and solar perturbations, exhibited in \cref{fig:SunMoon_1y}, is considerably more complex.
The quadrupolar approximation for the lunar perturbing function fails due to the large value of the semi-major axis (equal to approximately \num{0.7} times that of the Moon), and the averaged trends depart from the osculating.
Also, lunar encounters \review{at} the end of January, 1960 and in mid March, 1960, corresponding to \num{100} and \num{150} days after the initial epoch, respectively, impart sudden changes in the osculating orbital elements.
The outcome of the January approach is such that the post-encounter eccentricity is almost stationary, and it increases to the collisional value after the March flyby.
The signature of solar perturbations of intermediate period is recognizable in the behavior of the osculating inclination, which is above \ang{90} during significant periods of time before the January encounter.
We highlight that oscillations of intermediate period affect the geometry (and thus the outcome) of the March, 1960 encounter, and therefore the trajectory cannot be well modeled without considering either of these phenomena.

Ultimately, we cannot gain much insight on the osculating dynamics by looking only at the \review{(singly or doubly)} averaged \review{trajectories}, because these neglect the consequential effects of lunar close encounters on the trajectory evolution and on the lifetime.
We outline the implications for the Lidov-Kozai solution in more detail in the next section.

\section{Relationship with the Lidov-Kozai solution}
\label{sec:Luna3_LKE}
In \cref{sec:4bodyHamiltonian} we showed that the doubly-averaged perturbing function for the case in which two coplanar perturbers are present has the same form as that for a single perturber, except for a multiplicative constant that has the effect of changing the characteristic timescale.
Since the inclination of the Moon with respect to the ecliptic is small, the presence of two perturbers does not essentially preclude the existence of Lidov-Kozai solutions.
However, the doubly-averaged solution is obtained under the assumption that the semi-major axes ratios with respect to each of the perturbers are very small, and that orbits do not intersect; both of these assumptions fail in the case of Luna 3.
It is therefore natural to question whether its trajectory can still be categorized as a Lidov-Kozai solution.
The answer can be found by examining the values of the integrals $a, c_1, c_2$, \review{in addition to the perigee distance,} and by checking if the evolution can be predicted \review{through} Lidov-Kozai diagrams.

\subsection{Integrals of the doubly-averaged problem \review{and radius of perigee}}
\begin{figure}[t]
  \centering
  \begin{subfigure}[b]{0.5\linewidth}
    \centering
    \includegraphics[scale=0.85]{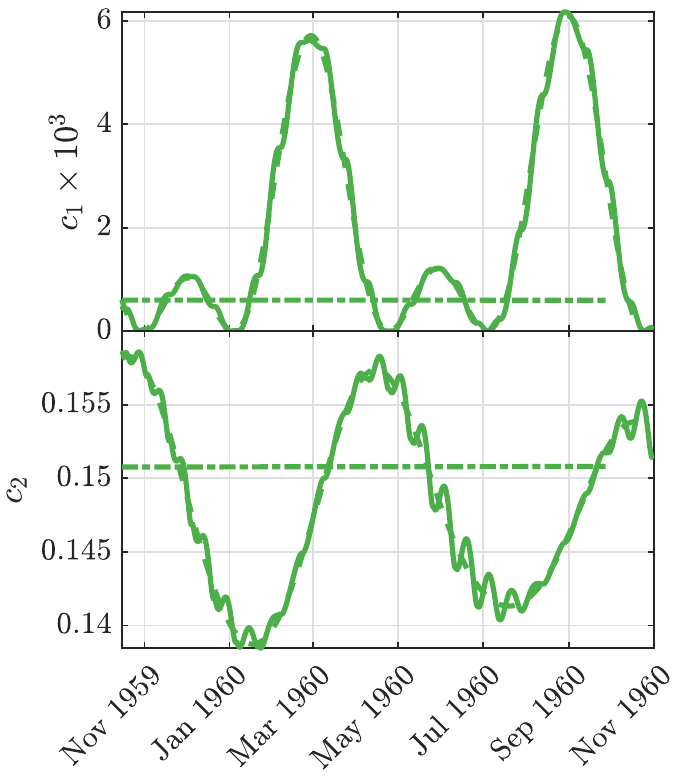}
    \caption{Sun only.\label{fig:c1_c2_S}}
  \end{subfigure}%
  \begin{subfigure}[b]{0.5\linewidth}
    \centering
    \includegraphics[scale=0.85]{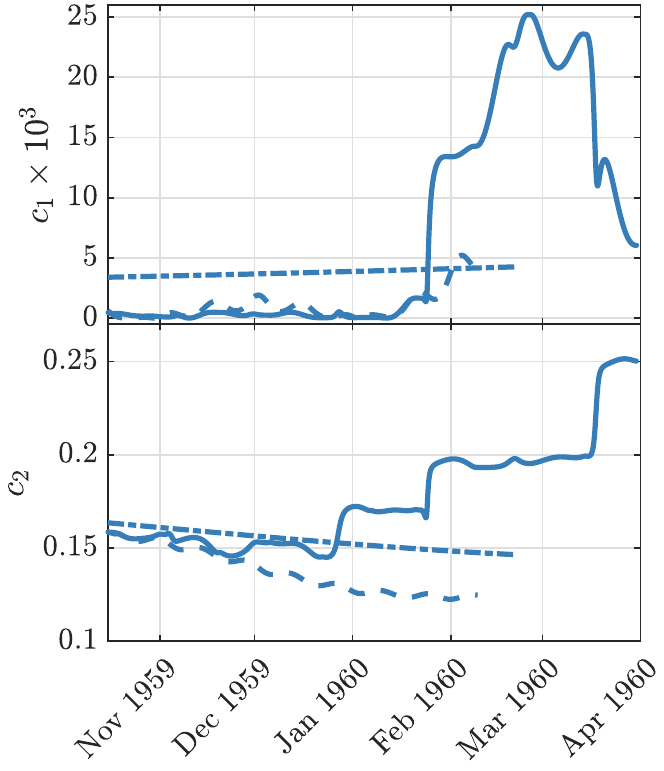}
    \caption{Sun, Moon.\label{fig:c1_c2_SM}}
  \end{subfigure}%
  \caption{Integrals of the doubly-averaged problem $c_1, c_2$ as a function of time obtained by plugging into \cref{eq:c1,eq:c2} osculating elements (continuous lines), singly-averaged elements (dashed lines), and doubly-averaged elements (dot-dashed lines).\label{fig:c1_c2}}
\end{figure}

\begin{figure}[tb]
  \centering
  \includegraphics[scale=0.8]{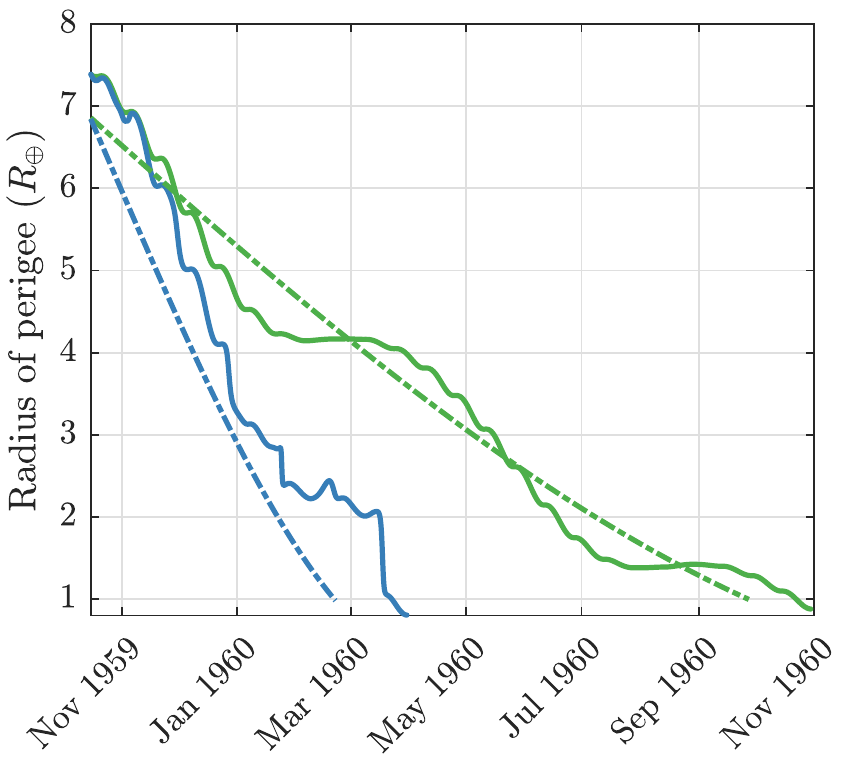}
  \caption{\review{Evolution of the radius of perigee starting from 15 October 1959 until collision with the Earth is detected. Continuous and dot-dashed lines denote osculating and doubly-averaged solutions, respectively. Either solar perturbations exclusively (in green) or both lunar and solar perturbations (in blue) are considered. The osculating solution with both lunar and solar perturbations is the same as in \cref{fig:rp}. The lifetime changes depending on the perturbations and on whether the trajectory is either singly- or doubly-averaged.\label{fig:rp_oscVsAvg}}}
\end{figure}

If osculating elements are plugged into \cref{eq:c1,eq:c2}, the quantities $c_1$ and $c_2$ will deviate from their doubly-averaged values (which are constant for small $a$).
If the osculating solution does not diverge from the doubly-averaged \review{trajectory}, deviations in $c_1, c_2$ are bounded.
Therefore, by evaluating $c_1, c_2$ on the osculating trajectory we can assess whether its dynamics can be well approximated by the Lidov-Kozai solution; the same procedure can also be used for the singly-averaged trajectory.

As shown in \cref{fig:c1_c2_S}, deviations of $c_1$ and $c_2$ are bounded if the Sun is the only perturber.
The quantity $c_1$, when evaluated on the osculating and singly-averaged solutions, has large excursions from the doubly-averaged value because of oscillations of intermediate period in inclination and eccentricity; nevertheless, it is clearly periodic and bounded.
All of the assumptions underlying the Lidov-Kozai theory are respected, and the dynamics are of the Lidov-Kozai type.
Conversely, if the Moon is taken into account as an additional perturber (\cref{fig:c1_c2_SM}), $c_1$ and $c_2$ depart from the doubly-averaged values due to lunar close encounters.
Therefore, predicting the evolution of the osculating trajectory through the Lidov-Kozai solution is bound to fail.
\review{As we consider the ecliptic inclination of the Moon to be nonzero for the averaged solutions ($i_\Moon = \ang{5;15;}$), we also observe small variations in the doubly-averaged $c_1, c_2$ in \cref{fig:c1_c2_SM}.}

\review{We compare the osculating and doubly-averaged histories of the radius of perigee in \cref{fig:rp_oscVsAvg}.
In the Sun-only case, in which the dynamics are well described by the Lidov-Kozai theory, the lifetime is of roughly 12 months.
This is twice the actual lifetime of the spacecraft, which is accurately predicted by the osculating solutions with lunisolar perturbations.
The presence of lunar close encounters extends the actual lifetime by a month with respect to that predicted by the doubly-averaged solution.}

\subsection{Lidov-Kozai diagrams}
\begin{figure}[t]
  \centering
  \begin{subfigure}[b]{.5\linewidth}
    \centering
    \includegraphics[width=8cm]{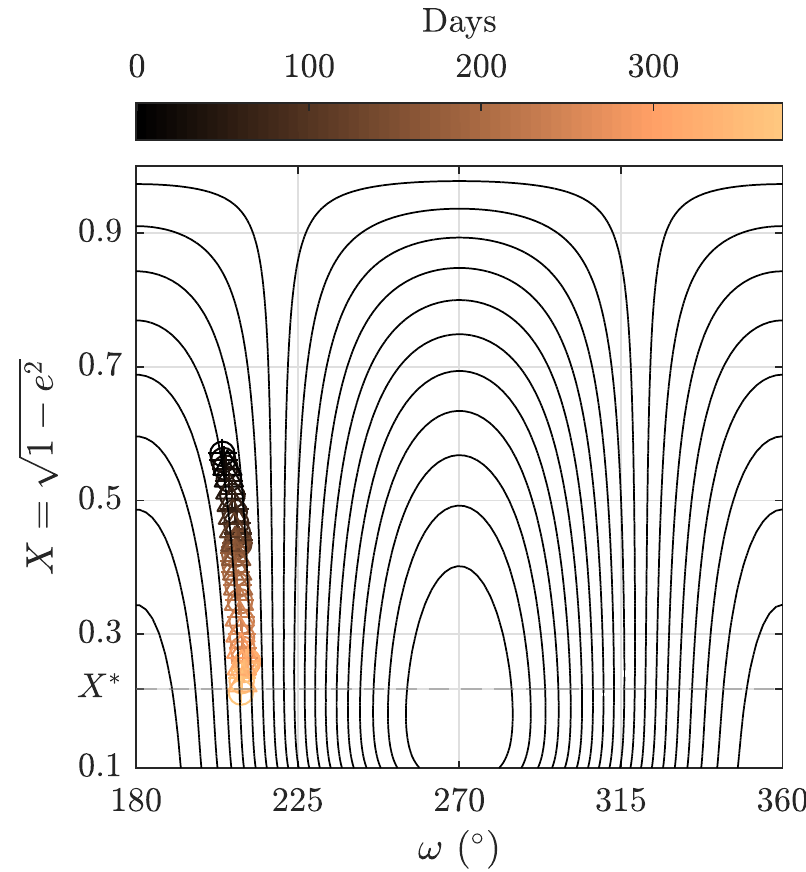}
    \caption{}
  \end{subfigure}%
  \begin{subfigure}[b]{.5\linewidth}
    \centering
    \includegraphics[width=8cm]{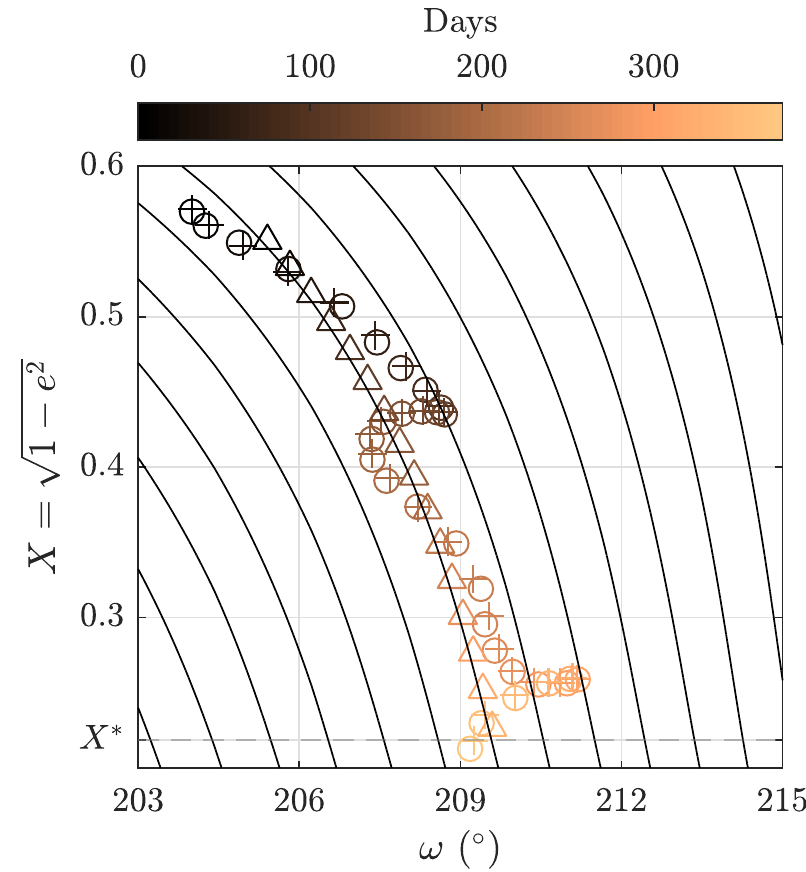}
    \caption{\label{fig:LK_S_Zoom}}
  \end{subfigure} 
  \caption{Lidov-Kozai diagram with solar perturbations only, with angles referred to the ecliptic. Left panel displays the $(\omega \geq \SI{180}{\degree}, X)$ half-plane, right panel is a zoom-in. Black lines are contours of the $c_2(\omega,e,c_1)$ integral, cfr. \cref{eq:c2_ec1w}. Circles, crosses, and \review{triangles} refer to results from the integration of non-averaged, singly-averaged, and doubly-averaged equations, respectively. The value $X^*$ corresponds to collision for the final value of osculating semi-major axis. The value of the integral $c_1$ is \num{5.98e-4}.  \label{fig:LK_S}}
\end{figure}

\begin{figure}[t]
  \centering
  \begin{subfigure}[b]{.5\textwidth}
    \centering
    \includegraphics[width=8cm]{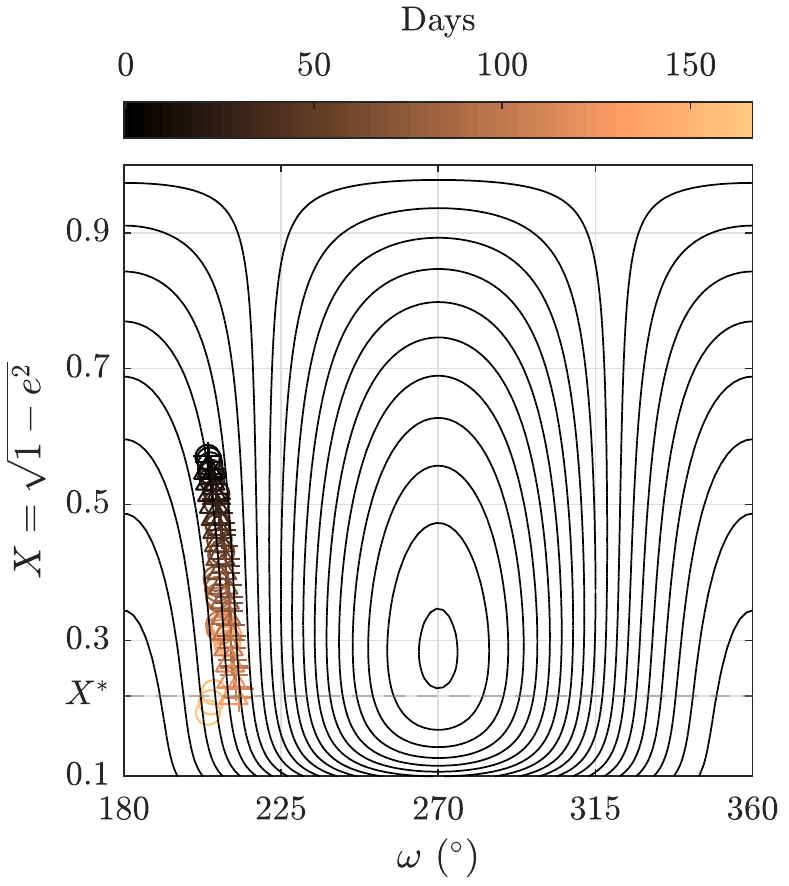}
    \caption{\label{fig:LK_SM_global}}
  \end{subfigure}%
  \begin{subfigure}[b]{.5\textwidth}
    \centering
    \includegraphics[width=8cm]{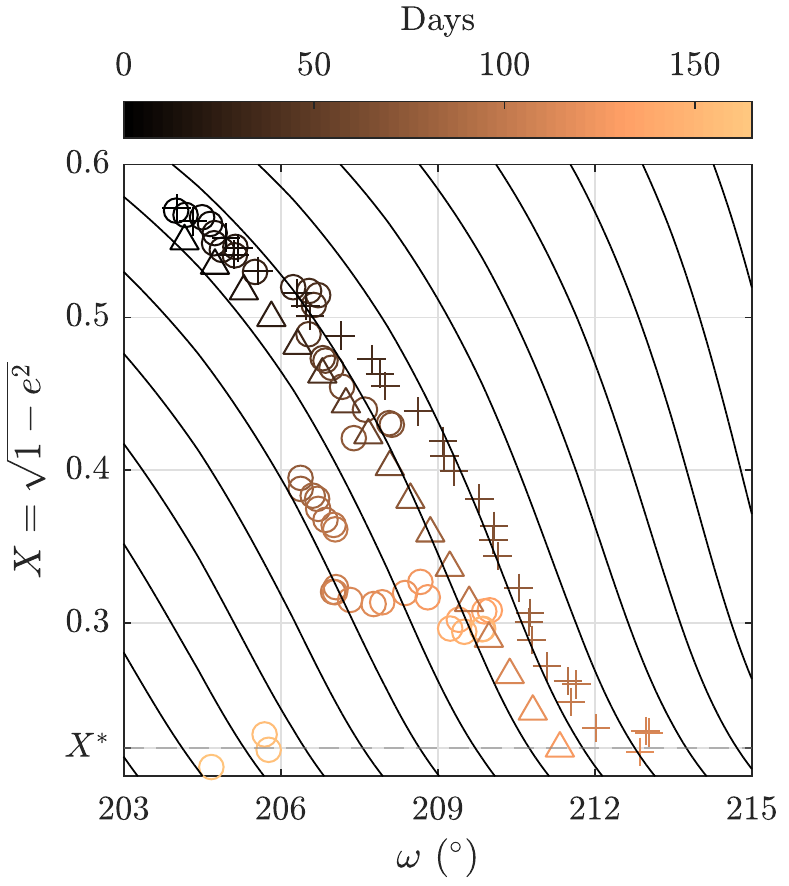}
    \caption{\label{fig:LK_SM_zoom}}
  \end{subfigure}
  \caption{Lidov-Kozai diagrams with both lunar and solar perturbations, with angles referred to the ecliptic. The description of the curves is as in \cref{fig:LK_S}. The value of the integral $c_1$ is \num{0.0038}.\label{fig:LK_SM}}
\end{figure}

We can get a qualitative idea of the modifications to the nature of the Lidov-Kozai solution induced by strong lunar perturbations by comparing trajectories with and without the Moon in Lidov-Kozai diagrams.
In \cref{fig:LK_S,fig:LK_SM}, we plot doubly-averaged, singly-averaged, and osculating trajectories on the diagrams simultaneously.
This is not a rigorous technique since the $c_2$-isolines only describe the doubly-averaged evolution; nevertheless, it allows us to efficiently highlight any differences in the secular behavior predicted by the doubly-averaged solution and that actually followed by the osculating \review{solution}.

As is evident from \cref{fig:LK_S_Zoom}, in the Sun-only case the trajectories closely follow the $c_2$-isolines and the secular trend of the osculating solution is well approximated by the doubly-averaged.
\review{Given that the orbit lifetime is less than the period of the perturber (which is of the same order of magnitude as the timescale of the intermediate-period perturbations), it is remarkable that the osculating evolution is not dominated by intermediate-period perturbations.}
Departures of the osculating and singly-averaged solution are due to the aforementioned intermediate-period terms in inclination and eccentricity.
Although introducing the Moon as a perturber does not change the global behavior of increasing eccentricity (\cref{fig:LK_SM_global}), a closer examination of the phase space reveals significant differences with respect to the Sun-only case.
As is visible from \cref{fig:LK_SM_zoom}, sudden changes in eccentricity due to lunar close encounters from \num{100} days onwards induce jumps of the osculating trajectory away from the evolution predicted by the doubly-averaged flow.

In essence, the trajectory is of the Lidov-Kozai type only if the Sun is considered as the exclusive perturber.
Adding the Moon results in the solution departing from the expected Lidov-Kozai behavior because of lunar close encounters.

\section{Conclusions}
\label{sec:concl}
Starting from the ephemerides published in the literature, we accurately reconstructed the cislunar trajectory of the Luna 3 spacecraft.
We confirm that the dynamics is completely described by a restricted Earth-spacecraft-Moon-Sun gravitational four-body problem, since additional perturbations ($J_2$ and higher harmonics, solar radiation pressure, and atmospheric drag) do not significantly change the trajectory.

Contrary to the commonly reported notion that Lidov-Kozai dynamics drove the spacecraft into colliding with the Earth, we find that its evolution was driven by a combination of oscillations of intermediate period induced by solar gravity and lunar close encounters.
Solar perturbations forced the eccentricity to acquire values corresponding to re-entry in a span of six months.
Such a short timescale, which is less than the period of the perturber in the Earth-Sun-spacecraft restricted problem, implies that intermediate period terms decisively shape the trajectory.
In particular, they are essential in reproducing the geometry of the March 1960 lunar close encounter, which provided a boost in eccentricity that further shortened the lifetime of the spacecraft.
At the quadrupolar order, solar perturbations of intermediate period also cause the singly-averaged and osculating inclinations to change quadrants, while the doubly-averaged inclination does not (in accordance with the classical Lidov-Kozai theory).



Ultimately, intermediate-period terms and lunar close encounters, both of which are neglected in the doubly-averaged solution, are as important as the secular trends of the doubly-averaged problem in determining the orbital evolution.
Impulsive perturbations engendered from lunar close encounters \review{break} the classical Lidov-Kozai dynamics because of sudden changes in the  $a, c_1, c_2$ integrals.
As a result, the post-encounter trajectory departs from the doubly-averaged flow.
These findings are of particular import for the trajectory design of cislunar and translunar missions, and for dynamical studies of exoplanetary systems.

As a final remark, this article treated the problem of large semi-major axis ratio, which invalidated the assumptions made by Lidov and Kozai in their quadrupole-level treatment as far as the Moon's perturbations are concerned.
It is well known that other dynamical effects can suppress Lidov-Kozai dynamics, the most common being the oblateness perturbation.
The Earth's oblateness and lunisolar perturbations approximately become equal at the Laplace radius of \num{7.7} Earth radii.
Beyond this distance, however, lunar dynamics of higher order become increasingly relevant, and thus it can be noted that in no region about Earth we can expect to have true Lidov-Kozai cycles.
Ironically, a phenomenon first discovered in the circumterrestrial domain has no physical application in this realm.
This may be the reason that Kozai's application to asteroidal dynamics gained widespread acclaim, whereas Lidov received only minor recognition until as of late.


\section*{Acknowledgements}
Parts of this work were presented at the 2018 John L. Junkins Dynamical Systems Symposium and at the 2019 Meeting of the AAS Division on Dynamical Astronomy (DDA).

Davide Amato thanks Jay McMahon for his indispensable support during the writing of this article, Giulio Baù for comments that improved the quality of the article, and Giovanni Valsecchi for helpful discussions at the 2019 DDA Meeting about averaged solutions in the presence of orbit crossings.

Renu Malhotra acknowledges funding from NSF (Grant AST-1824869), and the Marshall Foundation of Tucson, AZ, USA.

We acknowledge the use of software routines from the IAU SOFA Collection \citep{SOFA:2019-07-22} in the reduction of the Luna 3 ephemerides.

\bibliographystyle{spbasic}
\bibliography{main}

\end{document}